\documentclass[12pt]{extarticle}
\usepackage{setspace}

\usepackage[T1]{fontenc}

\usepackage[latin1]{inputenc}
\usepackage{epsfig}
\usepackage[english]{babel}
\usepackage{color}
\usepackage{graphicx}

\usepackage{dcolumn}

\usepackage{moreverb}

\usepackage{amsmath,amssymb,amsfonts}
\usepackage[all]{xy}

\begin{document}
\numberwithin{equation}{section}
\newcommand{\boxedeqn}[1]{%
  \[\fbox{%
      \addtolength{\linewidth}{-2\fboxsep}%
      \addtolength{\linewidth}{-2\fboxrule}%
      \begin{minipage}{\linewidth}%
      \begin{equation}#1\end{equation}%
      \end{minipage}%
    }\]%
}

%\boxedeqn{}

\newsavebox{\fmbox}
\newenvironment{fmpage}[1]
     {\begin{lrbox}{\fmbox}\begin{minipage}{#1}}
     {\end{minipage}\end{lrbox}\fbox{\usebox{\fmbox}}}

\raggedbottom
\onecolumn

\parindent 8pt
\parskip 10pt
\baselineskip 16pt
%\begin{flushleft}
\noindent\title*{{\LARGE{\textbf{Classical ladder operators, polynomial Poisson algebras and classification of superintegrable systems}}}}
\newline
\newline
Ian Marquette
\newline
School of Mathematics and Physics, The University of Queensland, Brisbane, QLD 4072, Australia
\newline
i.marquette@uq.edu.au
\newline
\newline
We recall results concerning one-dimensional classical and quantum systems with ladder operators. We obtain the most general one-dimensional classical systems respectively with a third and a fourth order ladder operators satisfying polynomial Heisenberg algebras. These systems are written in terms of the solutions of quartic and quintic equations. They are the classical equivalent of quantum systems involving the fourth and the fifth Painlev\'e transcendents We use these results to present two new families of superintegrable systems and examples of trajectories that are deformation of Lissajous's figures.

\section{Introduction}

The most well known system with ladder operators is the 1D harmonic oscillator with first order raising and lowering operators. Such operators first order creation and annihilation operators were also used in context of the N-dimensional isotropic and the anisotropic harmonic oscillator. The integrals of motion of these multi-dimensional systems can be written in terms of these operators [1,2,3]. The second order ladder operators of two-dimensional superintegrable Hamiltonians were obtained in the 60ties by Winternitz and al [4] and a classification of two-dimensional systems with second order ladder operators was performed later by Miller and Boyer [5]. More recently, systems with third order ladder operators were studied [6-19]. It was shown that a 1D quantum system with third order ladder operators can be constructed from supersymmetric quantum mechanics (SUSYQM) with a first and second order supercharges [11,16,19]. The potential, the supercharges, the ladder operators and the zero modes were written in terms of the fourth Painlev\'e transcendent ($P_{4}$). The case of quantum 1D systems with fourth order ladder operators and the relation with the fifth Painlev\'e transcendent was first explored in Ref.20 and Ref.21 and also later with extensions to 2D superintegrable systems and the relations with integrals of motion [22]. These systems were written explicitly in terms of the fifth Painlev\'e transcendent ($P_{5}$). These systems with higher order ladder operators can allow many zero modes for the annihilation and creation operators. It was discussed how superintegrable systems generated from such 1D systems can have very interesting spectrum with more complicated degeneracies [16,22] due to the presence of singlet, doublet and triplet states.

In classical mechanics, a particular case of a system with third order integral of motion [13,23] was related to third order ladder operators [24]. Let us mention that classical systems with third order ladder operators were also explored in an earlier paper of Eleonsky and Korolev [18]. The purpose of this paper is to present a classification of classical systems with third and fourth order ladder operators satisfying polynomial Heisenberg algebras and show how these results can be used to generate superintegrable systems.

The polynomial (i.e. polynomial in the momenta) ladder operators described in Ref.11,16,18 and 19 satisfy polynomial Heisenberg algebras. However, for many quantum systems we need to introduce nonpolynomial ladder operators and generalized Heisenberg algebras. For example, 1D systems with quadratic energy spectrum and their classical equivalent allow such nonpolynomial ladder operators [18,25-28]. The infinite well and the Morse potentials are examples of such systems. The existence of raising and lowering operators is also not restricted to systems with separation of variables in Cartesian coordinates [3] and Hamiltonians with a scalar potential. Ladder operators in context of systems with a monopole interaction were discussed recently in the case of generalized MICZ-Kepler [29] and generalized Kaluza-Klein monopole [30].

These ladder operators are also related to other approaches in quantum and classical mechanics. The relations between integrals, supersymmetric quantum mechanics and ladder operators were discussed by various authors [1-5,14,15,22,31-36]. These connections provide new methods to study know superintegrable systems, or even obtain new one and their corresponding polynomial algebra [14,15,22,36].

Let us present the organization of the paper. In Section 2, we recall results on 1D systems with ladder operators in quantum and classical mechanics. In Section 3, we obtain classical systems with third order ladder operators that satisfy a polynomial Heisenberg algebra. In Section 4, we discuss the case of fourth order ladder operators. In Section 5, we recall a method discussed in earlier paper to generate superintegrable systems in context of classical mechanics from analog of lowering and raising operators [14]. We obtain from results of Section 3 and Section 4 two new families of superintegrable systems and moreover we present examples of trajectories.

\section{Quantum and classical systems with ladder operators}

Let us consider the following 1D Hamiltonian of the form :

\begin{equation}
H=\frac{P^{2}}{2}+V(x),\quad P=-i\hbar \partial_{x},
\end{equation}
with a polynomial ladder operators of order n (i.e. polynomial in momenta),
\begin{equation}
A^{\mp}= \begin{cases}
\pm i P^{n}+f_{n-1}(x)P^{n-1}+...\pm i f_{2}(x)P^{2}+f_{1}(x)P\pm if_{0} & even ,\\
P^{n}\pm i f_{n-1}(x)P^{n-1}+...\pm i f_{2}(x)P^{2}+f_{1}(x)P\pm if_{0} & odd ,
\end{cases}
\end{equation}

that satisfy a polynomial Heisenberg algebra in quantum mechanics or the classical equivalent Poisson algebra in the context of classical mechanics. 

In quantum mechanics, a polynomial Heisenberg algebra satisfy the following algebraic relations :
\begin{equation}
[H,A^{\dagger}]=\omega A^{\dagger},\quad [H,A]=-\omega A ,
\end{equation}
\begin{equation}
A A^{\dagger}=Q(H+\omega),\quad A^{\dagger}A=Q(H), \quad [A,A^{\dagger}]=P(H),
\end{equation}
with $P(H)=Q(H+\omega)-Q(H)$. The polynomial Heisenberg algebra generated by $\{H,A,A^{\dagger}\}$ provides informations on the spectrum of $H$. We have for $Q(H)$ a $n$ th-order polynomial in $H$. The annihilation operator can allow at most $n$ zero modes (i.e. a state such $A\psi=0$). In each axis we can have at most $n$ infinite ladders by acting iteratively with the creation operator. The creation operator can also allow zero modes (i.e. a state such $A^{\dagger}\psi=0$) and we can have finite ladder operators. Such operators and their corresponding polynomial Heisenberg algebras are important in context of coherent states [37].
 
In classical mechanics we can also define analog of raising and lowering operators [2] ($A^{-}$ and $A^{+}$) and the corresponding polynomial Heisenberg algebra. We replace commutation relations by Poisson brackets. These operators satisfy the algebraic relations :
\begin{equation}
\{H,A^{+}\}_{p}=\omega A^{+},\quad \{H,A^{-}\}_{p}=-\omega A^{-}\quad ,
\end{equation}
\begin{equation}
\{A^{-},A^{+}\}_{p}=P(H),\quad A^{-}A^{+}=A^{+}A^{-}=Q(H),
\end{equation}
where $P(H)$ and $Q(H)$ are polynomials. They are related by $P(H)=-i\omega \frac{\partial Q(H)}{\partial H}$. Generalizations of the algebraic structure in classical and quantum were obtained by considering $\omega=\omega(H)$. Ladder operators can also be useful in context of 1D systems and they can be used to generate time-dependent integrals of motion :
\begin{equation}
Q^{\pm}=A^{\pm}e^{\mp i \omega t},
\end{equation}
\begin{equation}
\frac{dQ^{\pm}}{dt}=\{H,Q^{\pm}\}_{p}+\frac{\partial Q^{\pm}}{\partial t}=0.
\end{equation}
The constant value of this integrals given by Eq.(2.7) can be written in the following form :
\begin{equation}
q^{\pm}=c(E)e^{\pm i\theta_{0}}.
\end{equation}
We have $A^{\pm}$ as given by Eq.(2.2) and thus
\begin{equation}
A^{\pm}=c(E)e^{\pm(\theta_{0}+\omega t)},
\end{equation}
generate equations that can be used to obtain the trajectories on the phase space $(x(t),P(t))$ algebraically [27]. Ladder operators are thus useful in context of quantum mechanics but also classical mechanics.

\subsection{Classification of systems in classical and quantum mechanics with ladder operators}

The most general system with first and second order ladder operators in classical and quantum mechanics coincide [5] :

First order :
\begin{equation}
V_{1}=\frac{\omega^{2}}{2}x^{2},
\end{equation}

Second order :
\begin{equation}
V_{2}=\frac{\omega^{2}}{2}x^{2}+\frac{b}{x^{2}},
\end{equation}

These potentials also appear in context of superintegrable systems [4]. Recently systems with third and fourth order ladder operators were obtained and studied [11,16,21,22]. 

Third order :
\begin{equation}
V_{3}(x)=\frac{\omega_{1}^{2}}{2}x^{2}+\frac{\hbar\omega\epsilon}{2}P_{4}^{'}+\frac{\omega \hbar}{2}P_{4}^{2}+\omega \sqrt{\hbar \omega}xP_{4}+\frac{\hbar\omega}{3}(-\alpha+\epsilon) \quad ,
\end{equation}           
with  ($P_{4}=P_{4}(\sqrt{\frac{\omega}{\hbar}}x,\alpha,\beta)$ and $\epsilon=\pm 1$).

Fourth order :
\begin{equation}
V_{4}(x)=\frac{\omega^{2}}{8}(1+ \frac{4(P_{5}+\epsilon P_{5}')^{2}-5P_{5}^{2}}{(P_{5}-1)^{2}P_{5}})x^{2}+ \frac{\hbar^{2}}{x^{2}}(a-b-\frac{1}{8}+\frac{b-a P_{5}^{2}}{P_{5}}) -\hbar\omega(1+\frac{(2- \epsilon+2(1+c)P_{5})}{2(P_{5}-1)})
\end{equation}
with ($P_{5}=P_{5}(\frac{\omega}{\hbar}x^{2}, a,b,c,-\frac{1}{8})$ and $\epsilon=\pm 1$). The second order differential equations ( i.e. Painlev\'e equations) satisfied by the fourth and fifth Painlev\'e transcendents are the following : 
\begin{equation}
P_{4}^{''}(z) = \frac{P_{4}^{'2}(z)}{2P_{4}(z)} + \frac{3}{2}P_{4}^{3}(z) + 4zP_{4}^{2}(z) + 2(z^{2} -
\alpha)P_{4}(z) +  \frac{\beta}{P_{4}(z)},
\end{equation}
\begin{equation}
P_{5}''(z)=(\frac{1}{2P_{5}(z)}+\frac{1}{P_{5}(z)-1})P_{5}'(z)^{2}-\frac{1}{z}P_{5}'(z)+\frac{(P_{5}(z)-1)^{2}}{z^{2}}(\frac{aP_{5}^{2}(z)+b}{P_{5}(z)})
\end{equation}
\[+\frac{cP_{5}(z)}{z}+\frac{dP_{5}(z)(P_{5}(z)+1)}{P_{5}(z)-1}.\]

The Eq.(2.15) and (2.16) are two of the six Painlev\'e equations that serve to define the six Painlev\'e transcendents [38]. The fourth and the fifth Painlev\'e transcendents allow solutions in terms of rational functions or special functions [39]. These systems given the Eq.(2.13) and (2.14) are related with higher order supersymmetric quantum mechanics (SUSYQM) and we refer the reader to the ref.11, 16, 21 and 22 for more details concerning the explicit form of the supercharges, ladder operators, zero modes of the raising and lowering operators and the corresponding polynomial Heisenberg algebras. They possess respectively a polynomial Heisenberg algebra of order three and four. The energy of the zero modes are written  respectively in terms of the parameters of the fourth and fifth Painlev\'e transcendents. The corresponding superintegrable systems were discussed in Ref. 14, 16 and 22. The Painlev\'e transcendents appear in many context in physics and the relation with quantum systems with ladder operator and superintegrability is very interesting. The study of the classical equivalent of such system is thus important and is an unexplored subject.

In context of classical mechanics, the study of systems with higher order ladder operators (i.e. order higher than 2) is a relatively unexplored subject. System with third order ladder operators in classical mechanics were explored by Eleonsky and Korolev [18], and also in context of superintegrable systems [24]. Let us present the system considered in Ref.24 :
\begin{equation}
V(x)=\frac{\omega^{2}}{18}(2b + 5x^{2} + \epsilon 4x\sqrt{b+x^{2}}).
\end{equation}

A new family of superintegrable systems [21] were also constructed using the existence of ladder operators. We will present the general systems possess a third order ladder operators and classify systems with fourth order ladder operators. As for higher order superintegrability, the classical and quantum systems with higher order ladder operators differ.

\section{Classical systems with ladder operators of order 3}

From Eq.(2.2), a general third order raising and lowering operators has the form :
\begin{equation}
A_{x}^{\mp}=P_{x}^{3}\pm if_{2}(x)P_{x}^{2}+f_{1}(x)P_{x}\pm if_{0}(x).
\end{equation}

From Eq.(2.5) we obtain the following differential equations :
\begin{equation}
-i\omega + if_{2}^{'}(x)=0,
\end{equation}
\begin{equation}
-\omega f_{2}(x)-f_{1}^{'}(x)+3 V^{'}(x) =0,
\end{equation}
\begin{equation}
-i\omega f_{1}(x) +if_{0}^{'}(x)-2i f_{2}(x)V^{'}(x) =0,
\end{equation}
\begin{equation}
-\omega f_{0}(x) + f_{1}(x) V^{'}(x)=0 .
\end{equation}
The first two equations give us ( the terms that contain the two integration constants can be removed by an appropriate translations of $x$ and $V(x)$ ) :
\begin{equation}
f_{2}(x)=\omega x,\quad f_{1}(x)=-\frac{1}{2}\omega^{2} x^{2}+3V(x).
\end{equation}

When we put this in the last two equations we obtain :
\begin{equation}
-\omega f_{0}^{'}+V^{'}(x)(-\omega^{2}+3V^{'}(x))+(-\frac{1}{2}\omega^{2}x^{2}+3V(x))V^{'}(x)=0,
\end{equation}
\begin{equation}
f_{0}^{'}=\frac{1}{2}(-\omega^{3}x^{2}+6\omega V(x)+4\omega xV^{'}(x)).
\end{equation}

We can obtain a second order nonlinear differential equation for $V(x)$ :
\begin{equation}
\frac{\omega^{4}x^{2}}{2}-3\omega^{2} V(x) - 3 \omega^{2} x V^{'}(x) +3 (V^{'}(x))^{2}-\frac{1}{2}\omega^{2}x^{2} V^{''}(x)+3V(x)V^{''}(x).
\end{equation}

This equation coincide with the one obtained in Ref.18 (i.e. the Eq.5.3) and the one obtained in context of 2D superintegrability with second and third order integrals by Gravel [13] (more precisely the Eq.(19) in the limit $\hbar \rightarrow 0$). As shown in Ref.13 and also Ref.18, this equation can be integrated by using the fact that it possesses a first integral. 

However instead of giving the solution directly, let us take an other approach and impose the constraint that the ladder operators satisfy a polynomial Heisenberg algebra. The problem is reduced to 
solve a system of algebraic equations instead of a system of differential equations. Thus we consider the second relation of the Eq.(2.6). The polynomial $Q(H)$ can be at most a third order polynomial of the Hamiltonian with the following form :
\begin{equation}
A_{x}A_{x}^{+}=Q(H)=8(H-\epsilon_{1})(H-\epsilon_{2})(H-\epsilon_{3}).
\end{equation}

Instead of a set of differential equations as the one obtained from Eq.(2.5), the Eq.(3.10) generates a set of algebraic relations among $f_{0}(x)$, $f_{1}(x)$, $f_{2}(x)$ and $V(x)$
involving three parameters ( $\epsilon_{1}$, $\epsilon_{2}$ and $\epsilon_{3}$ ) :
\begin{equation}
2 \epsilon_{1}+2 \epsilon_{2}+2 \epsilon_{3}+2 f_{1}(x)+f_{2}^{2}(x)-6 V(x)=0,
\end{equation}
\begin{equation}
-4 \epsilon_{1} \epsilon_{2}-4 \epsilon_{1} \epsilon_{3}-4\epsilon_{2} \epsilon_{3}+f_{1}^{2}(x)+2 f_{0}(x) f_{2}(x)+8 \epsilon_{1} V(x)+8 \epsilon_{2} V(x)+8 \epsilon_{3} V(x)-12 V(x)^{2}=0,
\end{equation}
\begin{equation}
8 \epsilon_{1} \epsilon_{2} \epsilon_{3}+f_{0}^{2}(x) -8 \epsilon_{1} \epsilon_{2} V(x)-8 \epsilon_{1} \epsilon_{3} V(x)-8 \epsilon_{2} \epsilon_{3} V(x),
\end{equation}
\[ +8 \epsilon_{1} V(x)^{2}+8 \epsilon_{2} V(x)^{2}+8 \epsilon_{3} V(x)^{2}-8 V(x)^{3}=0.\]

We can use the expressions given by Eq.(3.6) for $f_{1}$ and $f_{2}$ and insert them in Eq.(3.11), (3.12) and (3.13). We obtain $\epsilon_{1}+\epsilon_{2}+\epsilon_{3}=0$ and the following two equations with 2 parameters $c$ and $d$ ( with $d=4(\epsilon_{1}^{2}+\epsilon_{1}\epsilon_{2}+\epsilon_{2}^{2})$ and $c=32 \omega^{2}(\epsilon_{1}^{2}\epsilon_{2}+\epsilon_{1}\epsilon_{2}^{2})$) :

\begin{equation}
\frac{\omega^{4} x^{4}}{4}+d+2 \omega x f_{0}(x)-3 \omega^{2} x^{2} V(x)-3 V(x)^{2}=0,
\end{equation}
\[ -\frac{1}{4\omega^{2}}c+f_{0}^{2}(x)+2d V(x)-8 V(x)^{3}=0 .\]

The solution is thus (including the Eq.(3.6)):
\begin{equation}
f_{0}(x)=\frac{3}{2\omega x}V^{2}(x) +\frac{3}{2}\omega x V(x)-\frac{1}{8}\omega^{3}x^{3}-\frac{d}{2\omega x},
\end{equation}
\begin{equation}
-9V^{4}(x)+14\omega^{2}x^{2}V^{3}(x)+(6d -\frac{15}{2}\omega^{4}x^{4})V^{2}(x)+ (\frac{3}{2}\omega^{6}x^{6}-2d)V(x)- (\frac{1}{2}d \omega^{4}x^{4}+d^{2})=0.
\end{equation} 
This solution satisfy all the equations given by Eq.(3.2)-(3.5) and (3.11)-(3.13). This solution agree with results obtained in Ref.18 and 13. The Eq.(3.14) coincide with Eq(5.5) of Ref.18 (up to a translation of the coordinates x and the potential. The Eq.(3.16) is a polynomial equation of degree 4 and thus can be solved. However, let us do not present explicitly this solution and only present a particular case. By imposing two of the roots of the polynomial to be equal (i.e. $\epsilon_{1}=\epsilon_{2}$) or ( $c^{2}=\frac{64d^{3}\omega^{4}}{27}$) we obtain the following solutions
\begin{equation}
f_{0}=-6\epsilon_{2} \omega x +\omega^{3} x^{3} ,\quad V(x) = -2 \epsilon_{2} +\frac{1}{2}\omega^{2}x^{2},
\end{equation}
and
\begin{equation}
f_{0}= 6\epsilon_{2}\omega x+ \frac{13}{27}\omega^{3}x^{3}\pm \frac{4\epsilon_{2}\sqrt{18\epsilon_{2}\omega^{2}x^{2}+\omega^{4}x^{4}}}{3 \omega x}\pm \frac{14}{27} \omega x \sqrt{18 \epsilon_{2} \omega^{2} x^{2}+\omega^{4} x^{4}},
\end{equation}
\[ V(x)=2 \epsilon_{2} + \frac{5 \omega^{2}x^{2}}{18} \pm \frac{2}{9} \sqrt{18 \epsilon_{2}x^{2}\omega^{2} +\omega^{4}x^{4}}.\]

We have the following relation $\epsilon_{2}=\sqrt{\frac{d}{12}}$. We can also use the parameter $d=\frac{\omega^{4}\tilde{d}}{27}$ to recover the potential obtained in Ref.13.

\section{systems with fourth order ladder operators}

Let us consider the ladder operators of the following form (i.e. as given by Eq.(2.2)) :
\begin{equation}
A_{x}^{\mp}=\pm iP^{4}+f_{3}(x)p^{3}\pm if_{2}(x)p^{2}+f_{1}(x)p\pm if_{0}(x).
\end{equation}

We obtain from Eq.(2.5) the following five differential equations :
\begin{equation}
-\omega-f_{3}^{'}(x)=0,
\end{equation}
\begin{equation}
-i\omega f_{3}(x)+if_{2}^{'}(x)-4iV^{'}(x)=0,
\end{equation}
\begin{equation}
-\omega f_{2}(x)-f_{1}'(x)+3f_{3}(x)V^{'}(x)=0,
\end{equation}
\begin{equation}
-i\omega f_{1}(x)+if_{0}^{'}(x)-2i f_{2}V^{'}(x)=0,
\end{equation}
\begin{equation}
-\omega f_{0}(x)+f_{1}(x)V^{'}(x)=0.
\end{equation}

From the Eq.(4.2) and (4.3) we obtain ( again by an appropriate translations of $x$ and $V(x)$ we can remove the terms involving the two integration constants ):
\begin{equation}
f_{2}=-\frac{1}{2}w^{2}x^{2}+4V(x),\quad f_{3}(x)=-\omega x.
\end{equation}

We insert the expressions given by Eq.(4.7) in the Eq.(4.4), (4.5) and (4.6). From the Eq.(4.6), we can isolate the functions $f_{0}(x)$ and $f_{0}^{'}(x)$ in terms of $f_{1}(x)$, $f_{1}^{'}(x)$, $V(x)$ and $V^{'}(x)$. Using these expresions, we obtain from the Eq.(4.4) and (4.5) :
\begin{equation}
-4\omega V(x)-f_{1}^{'}(x)+\frac{1}{2} \omega x (\omega^{2}x-6 V^{'}(x))=0,
\end{equation}
\begin{equation}
(\omega^{3}x^{2}-8\omega V(x)+f_{1}^{'}(x))V^{'}(x)+f_{1}(x)(-\omega^{2}+V^{''}(x))=0.
\end{equation}

From the Eq.(4.8) and the equation obtained by taking the derivative of the Eq.(4.8), we can isolate $f_{1}^{'}(x)$ and $f_{1}^{''}(x)$. We can thus insert these expresions for $f_{1}^{'}(x)$ and $f_{1}^{''}(x)$ in the equation obtained by isolating $f_{1}(x)$ in Eq.(4.9) and taking the derivative. We thus obtain for $V(x)$ the following third order nonlinear differential equation :
\begin{equation}
\omega^{4}x^{2}+2(-4\omega^{2}V(x)-6\omega^{2}xV^{'}(x)+(15V^{'}(x))^{2}+(16 V(x)+x(-2\omega^{2}x+9V^{'}(x)))V^{'}(x))
\end{equation}
\[ +\frac{3 V^{'}(x)(8V(x)+x(-\omega^{2}x+2V'(x)))V^{'''}(x)}{\omega^{2}-V''(x)}=0 ,\]

with $ \omega^{2}-V''(x) \neq 0 $. The solution $V''(x)=\omega^{2}$ is the well know harmonic oscillator.

Again, let us take the same approach than for systems with third order ladder operators and impose for the raising and lowering operators to generate a polynomial Heisenberg algebra (i.e. impose the Eq.(2.6) to be satisfied). The product of these opperators can be at most a quartic polynomial of the following form :

\begin{equation}
A_{x} A_{x}^{+}=16(H-\epsilon_{1})(H-\epsilon_{2})(H-\epsilon_{3})(H-\epsilon_{4}).
\end{equation}

Using the explicit form of the ladder operators given by Eq.(4.1), we obtain the following set of equations :
\begin{equation}
2(\epsilon_{1}+\epsilon_{2}+\epsilon_{3}+\epsilon_{4})+2 f_{2}(x)+f_{3}^{2}(x)-8V(x)=0,
\end{equation}
\begin{equation}
-4 (\epsilon_{1} \epsilon_{2}+ \epsilon_{1} \epsilon_{3}+ \epsilon_{2} \epsilon_{3}+ \epsilon_{1} \epsilon_{4}+ \epsilon_{2}\epsilon_{4}+ \epsilon_{3} \epsilon_{4})
\end{equation}
\[+2 f_{0}(x)+ f_{2}^{2}(x)+2 f_{1}(x) f_{3}(x) + 12 (\epsilon_{1} + \epsilon_{2} + \epsilon_{3} + \epsilon_{4}) V(x)-24 V(x)^{2}=0,\]
\begin{equation}
8 (\epsilon_{1} \epsilon_{2} \epsilon_{3}+ \epsilon_{1} \epsilon_{2} \epsilon_{4}+ \epsilon_{1} \epsilon_{3} \epsilon_{4}+ \epsilon_{2} \epsilon_{3} \epsilon_{4} )+f_{1}^{2}(x)+2 f_{0}(x) f_{2}(x)-16( \epsilon_{1} \epsilon_{2} + \epsilon_{1} \epsilon_{2} 
\end{equation}
\[+ \epsilon_{1} \epsilon_{3} + \epsilon_{2} \epsilon_{3} + \epsilon_{1} \epsilon_{4} + \epsilon_{2} \epsilon_{4} + \epsilon_{3} \epsilon_{4}) V(x)+24( \epsilon_{1}+ \epsilon_{2}+ \epsilon_{3} + \epsilon_{4}) V(x)^{2}-32 V(x)^{3}=0,\]
\begin{equation}
-16 \epsilon_{1} \epsilon_{2} \epsilon_{3} \epsilon_{4}+f_{0}(x)^{2}+16 (\epsilon_{1} \epsilon_{2} \epsilon_{3} + \epsilon_{1} \epsilon_{2} \epsilon_{4} + \epsilon_{1} \epsilon_{3} \epsilon_{4}+ \epsilon_{2} \epsilon_{3} \epsilon_{4} )V(x)
\end{equation}
\[-16 ( \epsilon_{1} \epsilon_{2} + \epsilon_{1} \epsilon_{3}+ \epsilon_{2} \epsilon_{3} + \epsilon_{1} \epsilon_{4} + \epsilon_{2} \epsilon_{4}+ \epsilon_{3} \epsilon_{4}) V(x)^{2}+16 ( \epsilon_{1} +\epsilon_{2} + \epsilon_{3} + \epsilon_{4}) V(x)^{3}-16 V(x)^{4}=0.\]

We will insert the expression given in Eq.(4.7) for $f_{3}(x)$ and $f_{2}(x)$ in Eq.(4.12)-(4.15). We obtain the following relation $\epsilon_{1}+\epsilon_{2}+\epsilon_{3}+\epsilon_{4}=0$ and using the following parameters :
\begin{equation}
d= \epsilon_{1}^{2}+\epsilon_{1} \epsilon_{2}+\epsilon_{2}^{2}+\epsilon_{1} \epsilon_{3}+\epsilon_{2} \epsilon_{3}+\epsilon_{3}^{2},
\end{equation}
\begin{equation}
 c= \epsilon_{1}^{2} \epsilon_{2}-\epsilon_{1} \epsilon_{2}^{2}-\epsilon_{1}^{2} \epsilon_{3}-2 \epsilon_{1} \epsilon_{2} \epsilon_{3}-\epsilon_{2}^{2} \epsilon_{3}-\epsilon_{1} \epsilon_{3}^{2}-\epsilon_{2}\epsilon_{3}^{2}, 
\end{equation}
\begin{equation} 
 e = \epsilon_{1}^{2} \epsilon_{2} \epsilon_{3}+\epsilon_{1} \epsilon_{2}^{2} \epsilon_{3}+\epsilon_{1} \epsilon_{2} \epsilon_{3}^{2}, 
\end{equation} 

we obtain three equations for $f_{0}(x)$, $f_{1}(x)$ and $V(x)$ :
\begin{equation}
-d+\frac{\omega^{4}x^{4}}{4}+2f_{0}(x)-2\omega x f_{1}(x)-4\omega^{2}x^{2}V(x)-8V^{2}(x)=0,
\end{equation}
\begin{equation}
8c-\omega^{2}x^{2}f_{0}(x)+f_{1}^{2}(x)-4 d V(x)+8 f_{0}(x)V(x)-32 V(x)^{3}=0, 
\end{equation}
\begin{equation} 
-16 e +f_{0}(x)^{2}+16 c V(x)-4 d V(x)^{2}-16 V^{4}(x)=0. 
\end{equation}

The solution is thus (including Eq.(4.7)) :
\begin{equation}
f_{0}(x)=\sqrt{-16e+16cV(x)-16dV(x)^{2}+16V(x)^{4}},
\end{equation}
\begin{equation} 
f_{1}(x)=\frac{1}{8\omega x}(16 d + \omega^{4} x^{4} +8 f_{0}(x)-16 \omega^{2}x^{2} V(x)-32 V^{2}(x)),
\end{equation}
\begin{equation}
\end{equation}
\[V(x)=-32 \omega^{2} x^{2} V^{5}(x)+ (128 d + 17 \omega^{4} x^{4}) V^{4}(x)+ (-96c-\frac{128d^{2}}{\omega^{2}x^{2}}-\frac{512e}{\omega^{2}x^{2}}-32d\omega^{2}x^{2}-\frac{7\omega^{6}x^{6}}{2})V^{3}(x)\]
\[+(-40d^{2}+352e+\frac{256c^{2}}{\omega^{4}x^{4}}+\frac{256cd}{\omega^{2}x^{2}}+32c\omega^{2}x^{2}+3d\omega^{4}x^{4}+\frac{11\omega^{8}x^{8}}{32})V^{2}(x)\]
\[+(-16cd-\frac{128cd^{2}}{\omega^{4}x^{4}}-\frac{512ce}{\omega^{4}x^{4}}-\frac{256c^{2}}{\omega^{2}x^{2}}+\frac{64d^{3}}{\omega^{2}x^{2}}+\frac{256de}{\omega^{2}x^{2}}-\frac{\omega^{10}x^{10}}{64})V(x)\]
\[+64c^{2}+4d^{3}-112de+\frac{16d^{4}}{\omega^{4}x^{4}}+\frac{128d^{2}e}{\omega^{4}x^{4}}+\frac{256e^{2}}{\omega^{4}x^{4}}-\frac{64cd^{2}}{\omega^{2}x^{2}}+\frac{256ce}{\omega^{2}x^{2}}-8cd\omega^{2}x^{2}\]
\[+\frac{3}{2}d^{2}\omega^{4}x^{4}+\frac{17}{2}e\omega^{4}x^{4}-\frac{1}{4}c\omega^{6}x^{6}+\frac{1}{64}d\omega^{8}x^{8}+\frac{\omega^{12}x^{12}}{4096}.\]
These relations satisfy well the Eq.(4.2)-(4.6) and Eq.(4.12)-4.16). Let us again do not present explicitly the solution for the potential that can be obtained because this is a quintic equation. We present a particular solution by considering a triple roots (i.e. $\epsilon_{1}=\epsilon_{2}$ and $\epsilon_{3}=\epsilon_{2}$ or
$c=\frac{2}{3}\sqrt{\frac{2}{3}}d^{\frac{3}{2}}$ and $e=\frac{d^{2}}{12}$).

We obtain the following solution :
\begin{equation}
f_{1}(x)=-\frac{(-8 \epsilon_{2}+ \omega^{2} x^{2})}{4 \omega x}, f_{0}(x)=-\frac{(-8\epsilon_{2}+\omega^{2}x^{2})^{3}((8\epsilon_{2}+\omega^{2}x^{2})}{16\omega^{4}x^{4}},
\end{equation}
\[ V(x)=-\epsilon_{2}+\frac{1}{8}\omega^{2}x^{2}+\frac{8\epsilon_{2}^{2}}{\omega^{2}x^{2}}, \]

and 

the following 1D system :
\begin{equation}
f_{0}(x)= -3 \epsilon_{2}^{2}+ \epsilon_{2} (-\frac{3}{2} w^{2} x^{2}\pm \frac{7}{16} \sqrt{64 \epsilon_{2} \omega^{2} x^{2}+\omega^{4} x^{4}})+\frac{1}{512}(-15 \omega^{4} x^{4}+17 \omega^{2} x^{2} \sqrt{64 \epsilon_{2} \omega^{2} x^{2}+
\omega^{4} x^{4}} ),
\end{equation}
\begin{equation}
f_{1}(x) = -6 \epsilon_{2} \omega x-\frac{3 \omega^{3} x^{3}}{32}\pm \frac{\epsilon_{2} \sqrt{64 \epsilon_{2} \omega^{2} x^{2}+\omega^{4} x^{4}}}{\omega x}\pm \frac{5}{32} \omega x \sqrt{64 \epsilon_{2} \omega^{2} x^{2}+\omega^{4} x^{4}},
\end{equation}

\begin{equation}
V(x) = \frac{1}{64} (96 \epsilon_{2}+5 \omega^{2} x^{2} \mp 3 \sqrt{64 \epsilon_{2} \omega^{2} x^{2}+\omega^{4} x^{4}}) .
\end{equation}

\section{New superintegrable systems in classical mechanics}

Let us now consider a classical 2D Hamiltonian separable in Cartesian coordinates
\begin{equation}
H(x_{1},x_{2},P_{1},P_{2})=H_{1}(x_{1},P_{1})+H_{2}(x_{2},P_{2}),
\end{equation}
for which polynomial ladder operators ($A_{x_{i}}$ and $A_{x_{i}}^{+}$) of order $n_{i}$ exist and satisfy polynomial Heisenberg algebras. From the Eq.(2.2) the operators $f_{1}=(A_{x_{1}}^{+})^{m_{1}}(A_{x_{2}}^{-})^{m_{2}}$ and $f_{2}=(A_{x_{1}}^{-})^{m_{1}}(A_{x_{2}}^{+})^{m_{2}}$ ( of order $n_{1}m_{1}+n_{2}m_{2}$) Poisson commute with the Hamiltonian H given by Eq.(2.1) if $m_{1}\omega_{x_{1}}=m_{2}\omega_{x_{2}}=\omega$ with $m_{1}$,$m_{2}$ $\in \mathbb{Z}^{+}$. The following products are also polynomial (i.e. polynomial in the momenta) integrals of the Hamiltonian H :
\begin{equation}
I_{1}=(A_{x_{1}}^{+})^{ m_{1}}(A_{x_{2}}^{-})^{ m_{2}}- (A_{x_{1}}^{-})^{ m_{1}}(A_{x_{2}}^{+})^{ m_{2}}, \quad I_{2}=(A_{x_{1}}^{+})^{ m_{1}}(A_{x_{2}}^{-})^{ m_{2}}+ (A_{x_{1}}^{-})^{ m_{1}}(A_{x_{2}}^{+})^{ m_{2}}\quad .
\end{equation}
By construction the Hamiltonian has the following second order integral related to separation of variables in Cartesian coordinates :
\begin{equation}
K=H_{1}-H_{2}.
\end{equation}
The Hamiltonian H given by Eq.(5.1) is thus maximally superintegrable (i.e. there are 3 functionally independent integrals of motion including the Hamiltonian himself). We construct from integrals given by Eq.(5.2) and (5.3) the following polynomial Poisson algebra :
\begin{equation}
\{K,I_{1}\}_{p}=2\omega I_{2},\quad \{K,I_{2}\}_{p}=2\omega I_{1},\quad \{I_{1},I_{2}\}_{p}=2 Q_{1}(\frac{1}{2}(H+K))^{m_{1}-1}
\end{equation}
\[Q_{2}(\frac{1}{2}(H-K))^{m_{2}-1}[m_{2}^{2}Q_{1}(\frac{1}{2}(H+K))P_{2}(\frac{1}{2}(H-K))-m_{1}^{2}Q_{2}(\frac{1}{2}(H-K))P_{1}(\frac{1}{2}(H+K))]. \]

\subsection{Two new families of superintegrable systems}
 
We can consider a new following 2D superintegrable system $H$ as given by Eq.(5.1) with $H_{1}$ and $H_{2}$ given by :

\begin{equation}
H_{1}=\frac{P_{x_{1}}^{2}}{2}+V_{1}(x_{1};\omega_{1}),\quad H_{2}=\frac{P_{x_{2}}^{2}}{2}+V_{2}(x_{2};\omega_{2}),
\end{equation}
 
with  $V_{1}(x_{1};\omega_{1})$ and $V_{2}(x_{2};\omega_{2})$ that both satisfy Eq.(3.16). We impose the constraint $m_{1}\omega_{x_{1}}-m_{2}\omega_{x_{2}}=0$ with $m_{1}$,$m_{2}$ $\in \mathbb{Z}^{+}$. From the construction at the beginning 
of the Section 5, this systems is thus maximally superintegrable.

Using the same form (i.e. the one given by Eq.(5.5)), we impose that $V_{1}$ and $V_{2}$ satisfy now the Eq.(4.24). We can thus generate a second family of 2D superintegrable systems. These results show how new systems with ladder operators can generate new superintegrable Hamiltonian. These results can be extended in N dimensions.
 
Let us present trajectories for a particular case (i.e. satisfying Eq.(4.28)) respectively for $\epsilon=1$ and $\epsilon=-1$. All bounded trajectories are closed and the motion is periodic. We choose for Fig.1 and Fig.2 the following parameters $\omega_{1}=1$, $\omega_{2}=2$, $\epsilon_{2_{x_{1}}}=4$, $\epsilon_{2_{x_{2}}}=16$, $x(0)=y(0)=1$, $x'(0)=1$, $y'(0)=-3$ and $t =[0,20]$.

\begin{figure}
 \begin{minipage}[b]{.45\linewidth}
  \centering\epsfig{figure=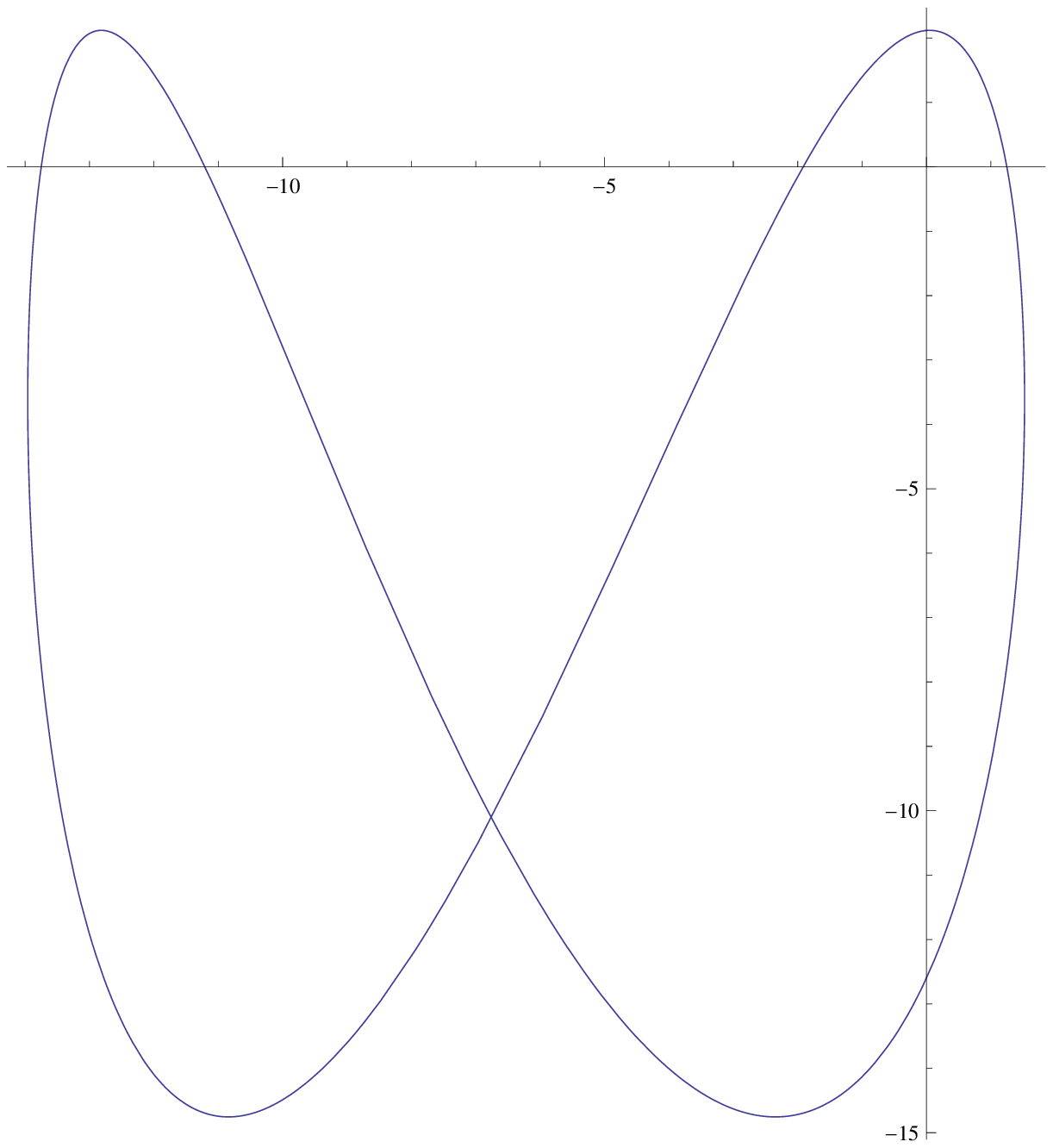,width=\linewidth}
% \unnumberedcaption{Fig. 1}
\caption{} 
 \end{minipage} \hfill
 \begin{minipage}[b]{.45\linewidth}
  \centering\epsfig{figure=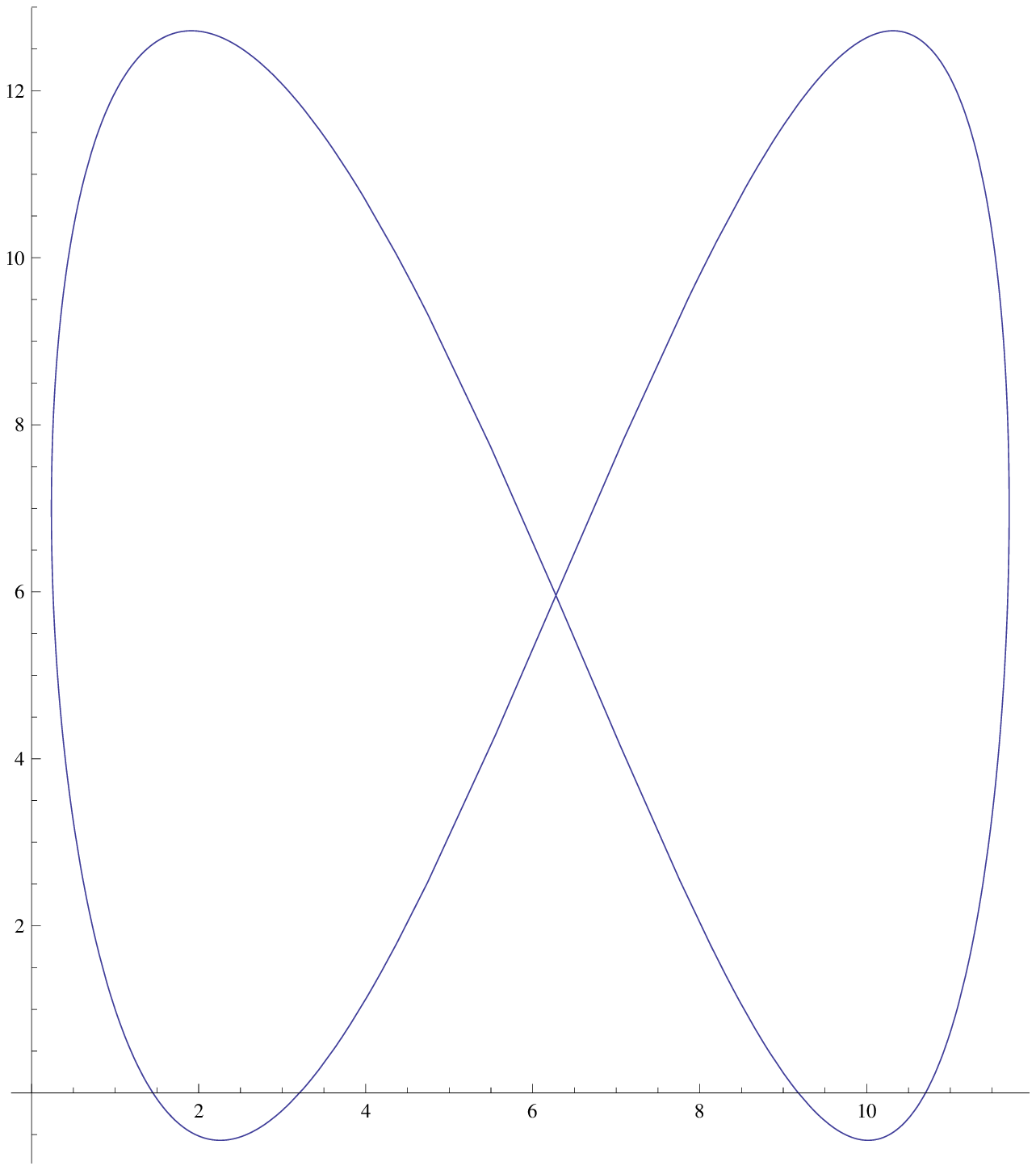,width=\linewidth}
%  \unnumberedcaption{Fig. 2}
\caption{}
 \end{minipage}
\end{figure}

\section{Conclusion}

We presented the most general 1D classical system with a third order ladder operators and also the most general system with fourth order ladder operators. We obtain respectively potentials that satisfy a fourth and a fifth order algebraic equations. They are the classical equivalent of the quantum systems written in terms of the fourth and fifth Painlev\'e transcendents obtained and studied in earlier papers [11,16,19,21,22]. This point out how systems with higher order ladder operators differ in classical and quantum mechanics.

Using a method introduced in earlier articles [14,24] and extended by Miller, Kalnins and Kress [36] to systems with separation of variables in polar coordinates, we obtained two new families of superintegrable systems that are deformations of the anisotropic harmonic oscillator and Evans-Verrier systems [40]. This paper point out also how ladder operators can be used to generate new superintegrable systems. Let us mention other recent papers that discuss ladder operators in context of superintegrability [41,42].

\textbf{Acknowledgments} The research of I.M. was supported by a postdoctoral research fellowship from FQRNT of Quebec. referees for their comments
and pointing out references.

%\LastPageEnding

\end{document}